\documentclass[11pt]{article}
\usepackage{moriond2,epsfig}

\def \BR#1  {\mbox{${\mathcal B}$(#1)} \ }
\def \dzero {\mbox{D\O}}
\def \gev   {\mbox{$GeV$}}
\def \gevoc {\mbox{$GeV/c$}}
\def \met   {\mbox{${\hbox{$E$\kern-0.6em\lower-.1ex\hbox{/}}}_T$}} 
\def \pb    {\mbox{$pb^{-1}$}}
\def \pbar  {\overline{p}}
\def \qbar  {\overline{q}}
\def \tev   {\mbox{$TeV$}}
\def \ra    {\rightarrow}
\def \be    {\begin{equation}}
\def \ee    {\end{equation}}
\def \bea   {\begin{eqnarray}}
\def \eea   {\end{eqnarray}}

\begin{document}
\vspace*{4cm}
\title{W/Z PRODUCTION CROSS SECTIONS AND ASYMMETRIES AT E$_{CM}$ = 2$\,TEV$}

\author{A. Bellavance for the \dzero\ and CDF collaborations}

\address{Department of Physics \& Astromony,
  University of Nebraska at Lincoln,\\
  116 Brace Laboratory,
  Lincoln, NE, 68588-0111, USA, \\
  bellavan@fnal.gov}

\maketitle
\abstracts{
The most recent results for W and Z boson production cross sections and
asymmetries are presented from the CDF and \dzero\ collaborations using 
Run II data taken at the Fermi National 
Accelerator Laboritory (FNAL) Tevatron. Data set sizes range 
from 72\pb to 226\pb, and results range from published to preliminary. 
Results presented agree with the Standard Model and world averages 
within errors.
}

\section{Motivation}\label{sec:motivation}

Decays of W and Z bosons into leptons provide some of the cleanest and simplest
processes with which to study electroweak interactions in the 
Standard Model (SM).
They provide: data with which one can gain a better understanding of 
the resolutions and efficiencies of one's detector and triggers; cross sections
to which other processes can be normalized; and
experimental tests of the validity of some SM parameters.

\section{\dzero\ and CDF Detectors and Analysis Methods}\label{sec:general}

CDF and \dzero\ are experiments studying
proton-antiproton collisions at a center of mass energy ($E_{CM}$) 
of 1.96\tev\  (``Run II''). Both detectors have a cylindrical geometry around
a proton-antiproton interaction region, and both have undergone
extensive upgrades since Run I data was collected at $E_{CM}$ of 1.8\tev.
Upgrade details have been published\cite{LeCompte:2000st}.


These analyses focus on lepton decay channels. The Z boson
events are required to have two energetic leptons. The W boson events are
required to have one energetic lepton, and missing transverse energy (\met)
as evidence of the undetected neutrino.
The majority of the backgrounds are from QCD processes, 
with levels estimated using QCD dijet data. 
Main systematic uncertainties are from parton distribution
functions (PDFs) (1-2\%).
Luminosity measurement uncertainties are about 6\%, and are
not considered as part of the other systematic uncertainties.

Electrons are required to have an isolated electromagnetic (EM) calorimeter 
cluster with a matching track. For muons, an isolated 
track is required that matches to an isolated calorimeter MIP or muon 
system track segment, and an appropriate timing coincidence and impact 
parameter are required.
For the tau channel, one or three isolated tracks plus a narrow jet and
reconstructible neutral pions are required. Hadronic decay 
products that reconstruct to the
tau mass are preferred through cuts or neural network parameters.

\section{Z Boson Cross Sections}\label{sec:Zxsections}
\subsection{$Z \ra ee$ Cross Section}\label{subsec:Zee}

To select $Z \ra ee$ events, both experiments require two
EM objects with transverse energies ($E_{T}$) 
greater than 25\gev.
The pseudorapidity ($\eta$) range selected by CDF for this study goes out to 
$\pm$2.8, which includes both the central and plug calorimeters. 
The results from \dzero\
have an $\eta$ range of $\pm$1.05, which includes only the 
central calorimeter. The largest background (about 2\%) comes from QCD 
dijet events. Dominant 
systematic uncertainties for this decay channel come from PDFs and electron
identification, each at about 1.5\%.

\begin{figure}[ht]
\begin{center}
\psfig{figure=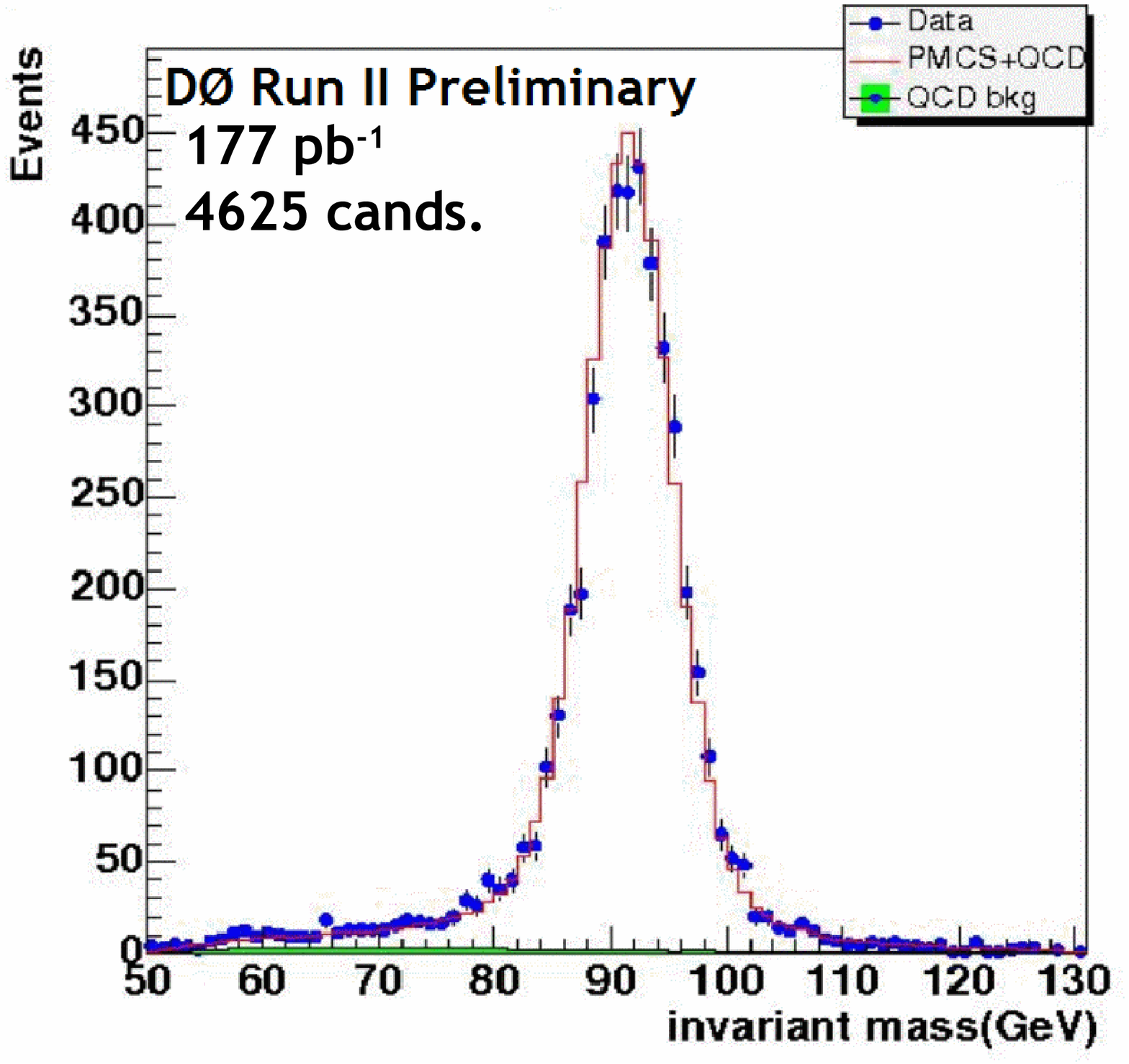,width=4.6in}
\end{center}
\caption{
The two electron invariant mass of $Z \ra ee$ events from
177\pb of \dzero\ data\protect\cite{D0:WZe}, with $|\eta|$ $<$ 1.05.
}
\label{fig:d0:Zee}
\end{figure}

\dzero\ has released preliminary results for a 177\pb data set\cite{D0:WZe}, 
and the resulting cross section is given in (\ref{eq:d0:Zee}). 
The two electron 
invariant mass for this data set is shown in Fig.~\ref{fig:d0:Zee}. 
CDF has published results for 72\pb of data\cite{CDF:WZe}, and the resulting 
cross section is given in (\ref{eq:CDF:Zee}). 
\bea
\dzero: \sigma \times \BR{$Z \ra ee$}\, =\,
264.9 \pm 3.9_{stat} \pm 9.9_{sys} \pm 17.2_{lum}\,pb
\label{eq:d0:Zee} \\
CDF: \sigma \times \BR{$Z \ra ee$}\, =\,
255.8 \pm 3.9_{stat} \pm 5.5_{sys} \pm 15_{lum}\,pb
\label{eq:CDF:Zee}
\eea

\subsection{$Z \ra \mu \mu$ Cross Section}\label{subsec:Zmumu}

$Z \ra \mu \mu$ events are selected by requiring two muon
objects with a minimum transverse momentum ($p_{T}$). For \dzero\ the cut was
$p_{T}$ $>$ 15\gevoc\ and for CDF the cut was $p_{T}$ $>$ 20\gevoc. 
Backgrounds are at
the 1\% level and include QCD events and $Z \ra \tau \tau$
decays. The largest systematic uncertainties for \dzero\ come from PDFs
and a Drell-Yan correction, each at about 1.5\%. The CDF analysis does
not apply a Drell-Yan correction, but does include PDF uncertainties.

\begin{figure}[ht]
\begin{center}
\psfig{figure=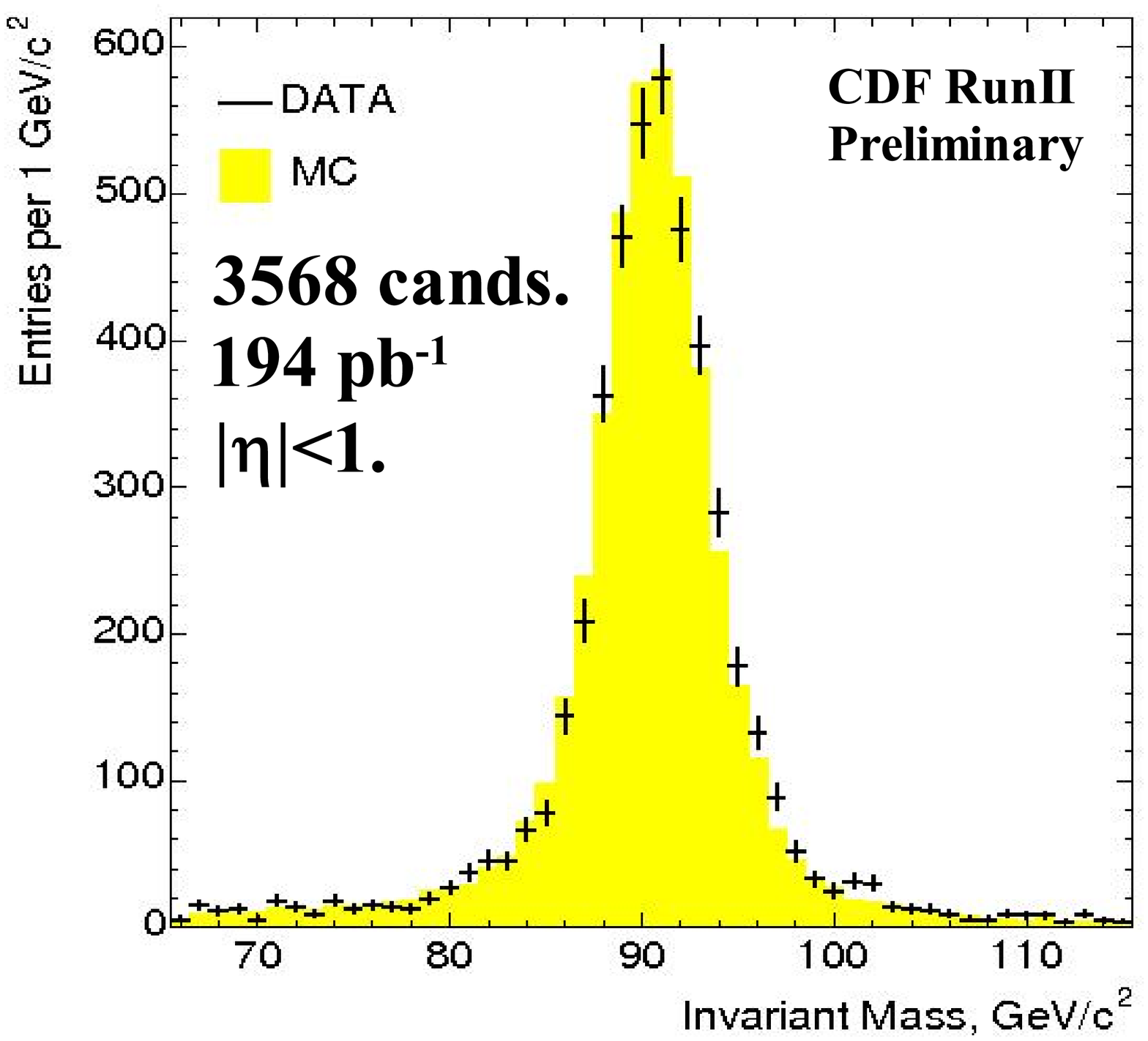,width=4.6in}
\end{center}
\caption{
The two muon invariant mass of $Z \ra \mu \mu$ events from 
194\pb of CDF data\protect\cite{Varganov:2004xr}, 
with $|\eta|$ $<$ 1.
}
\label{fig:cdf:Zmumu}
\end{figure}

\begin{figure}[ht]
\begin{center}
\psfig{figure=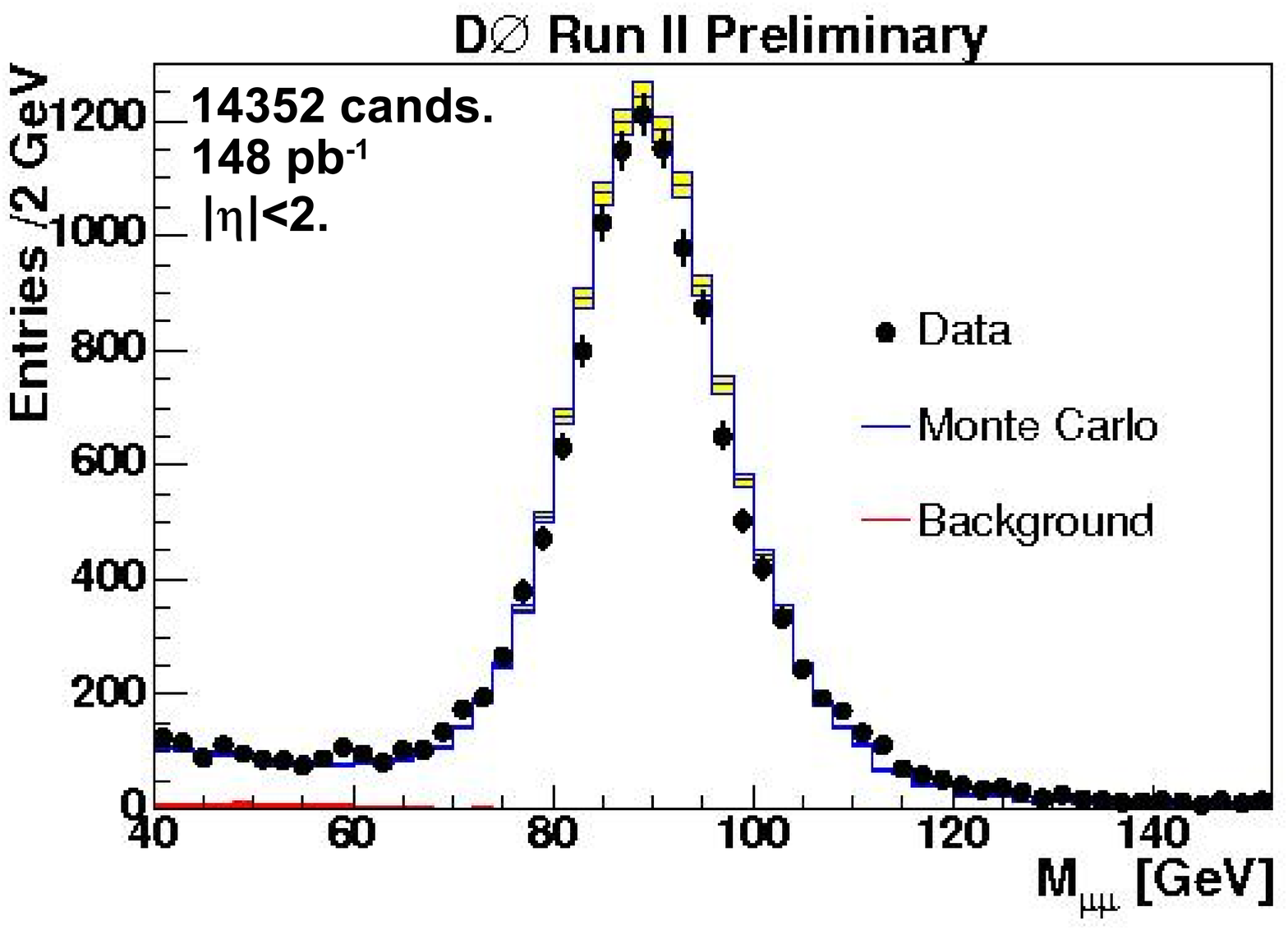,width=5.2in}
\end{center}
\caption{
The two muon invariant mass of $Z \ra \mu \mu$ events from 
148\pb of \dzero\ data\protect\cite{D0:Zmumu}, 
with $|\eta|$ $<$ 2.
}
\label{fig:d0:Zmumu}
\end{figure}

Both the \dzero\ and the CDF most recent results for this decay channel are
preliminary, using data sets of 148\pb and 194\pb, 
respectively\cite{D0:Zmumu,Varganov:2004xr}.
The cross sections are
given in (\ref{eq:d0:Zmm}) and (\ref{eq:CDF:Zmm}) and plots of the invariant 
two-muon reconstructed mass are shown in Figs.~\ref{fig:cdf:Zmumu} and
\ref{fig:d0:Zmumu}.
\bea
\dzero: \sigma \times \BR{$Z \ra \mu \mu$}\, =\,
291.3 \pm 3.0_{stat} \pm 6.9_{sys} \pm 18.9_{lum}\, pb
\label{eq:d0:Zmm} \\
CDF: \sigma \times \BR{$p\pbar \ra Z/\gamma^{*} \ra \mu \mu$}\, =\,
253.1 \pm 4.2_{stat} (^{+8.3})_{sys} \hspace{-1.25cm} _{-6.4} \hspace{0.6cm}
\pm 15.2_{lum} pb
\label{eq:CDF:Zmm}
\eea

\subsection{$Z \ra \tau \tau$ Cross Section}\label{subsec:Ztautau}

To distinguish taus from other leptons, both CDF and
\dzero\ require that one tau decay leptonically (into $e\nu_{e}\nu_{\tau}$
or $\mu\nu_{\mu}\nu_{\tau}$), and the other decay hadronically (into 
(1 or 3)$\pi^{\pm}\nu_{\tau}\,+\,N\pi^{0}$) for $Z \ra \tau \tau$ events. 
\dzero\ also requires that the taus have opposite charges. The largest 
backgrounds for this decay channel are dijet events 
(about 10\%) and other leptonic
Z decays (about 6\%).

\begin{figure}[ht]
\begin{center}
\psfig{figure=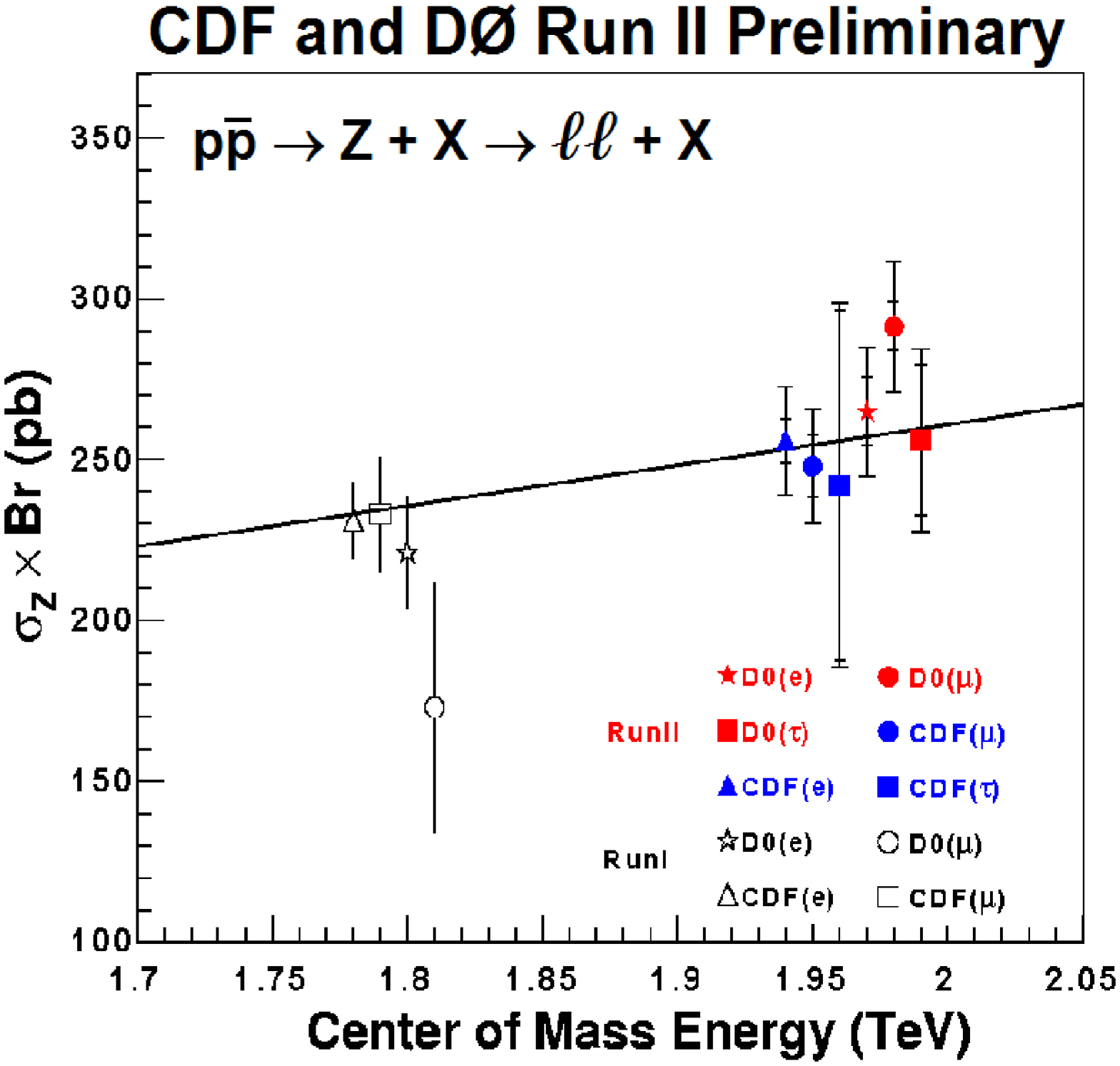,width=4.6in}
\end{center}
\caption{
Summary of Z boson
cross section measurements\protect\cite{Zsum}
from CDF and \dzero. Data was taken at 1.8\tev\ or
1.96\tev\ - points are spread for ease of reading.
The line is a Standard Model
prediction\protect\cite{Hamberg:1990np}.
}
\label{fig:Zxsec}
\end{figure}

Cross sections for $Z \ra \tau \tau$ are given in (\ref{eq:d0:Ztt}) and
(\ref{eq:CDF:Ztt}) for 226$pb^{-1}$ and 72$pb^{-1}$ of data, respectively. 
Details
are available in publications for both \dzero \cite{Abazov:2004vd}
and CDF\cite{Safonov:2004zv}.
Leptonic Z cross section results are compared in Fig.~\ref{fig:Zxsec}.
\bea
\dzero: \sigma \times \BR{$Z \ra \tau \tau$}\, =\,
237 \pm 15_{stat} \pm 18_{sys} \pm 15_{lum} pb
\label{eq:d0:Ztt} \\
CDF: \sigma \times \BR{$p\overline{p} \ra Z/\gamma^{*}
                       \ra \tau \tau$}\, =\,
242 \pm 48_{stat} \pm 26{sys} \pm 15_{lum} pb
\label{eq:CDF:Ztt}
\eea

\section{W Boson Cross Sections}\label{sec:Wxsections}
\subsection{$W \ra e \nu$ Cross Section}\label{subsec:Wenu}

To identify $W \ra e \nu$ decays, both experiments required an electron with
$E_{T}$ greater than 25\gev\ that matched to a track. \dzero\ loosened
this requirement to 20\gev\ for electrons in the central region of their
detector.
\met\ of more
than 25\gev\ is also required. The most significant backgrounds are
QCD dijet events and $Z \ra ee$ events at about 2\% each. The largest
uncertainties come from PDFs and electron identification at about 1.5\%
each.

CDF has results for a 72\pb data set\cite{CDF:WZe} 
that gives the cross section 
in (\ref{eq:CDF:Wen_c}) for the central part of their detector.
\dzero\ finds the cross section given in (\ref{eq:d0:Wen}) for 177\pb
of data\cite{D0:WZe}.
\bea
CDF: \sigma \times \BR{$W \ra e \nu$}\, =\,
2780 \pm 14_{stat} \pm 60_{sys} \pm 166_{lum} pb
\label{eq:CDF:Wen_c} \\
\dzero: \sigma \times \BR{$W \ra e \nu$}\, =\,
2865 \pm 8.3_{stat} \pm 76_{sys} \pm 186_{lum} pb
\label{eq:d0:Wen}
\eea

\subsection{$W \ra \mu \nu$ Cross Section}\label{subsec:Wmunu}

\dzero\ and CDF both require $W \ra \mu \nu$ events to have a muon track with
$p_{T}$ greater than 20\gev\ and \met\ greater
than 20\gev. The largest backgrounds come from similar types of decays 
($Z \ra \mu \mu$ and $W \ra \tau \nu$ at about 6\%), and from QCD b-jets
(about 1\% as calculated from data). The largest systematic uncertainties
come from efficiencies (about 1.5\%) and PDFs (about 1\%).

CDF has a result for 194\pb of data\cite{Varganov:2004xr} that 
results in the cross section given in (\ref{eq:CDF:Wmn}). \dzero\ has
a preliminary cross section given in (\ref{eq:d0:Wmn}) for 96\pb of
data\cite{D0:Wmunu}.
\bea
CDF: \sigma \times \BR{$W \ra \mu \nu$}\, =\,
2786 \pm 12_{stat} \pm (^{+65})_{sys} \hspace{-1.15cm} _{-55} \hspace{0.6cm}
\pm 166_{lum} pb
\label{eq:CDF:Wmn} \\
\dzero: \sigma \times \BR{$W \ra \mu \nu$}\, =\,
2989 \pm 15_{stat} \pm 81_{sys} \pm 194_{lum} pb
\label{eq:d0:Wmn}
\eea
\begin{figure}[ht]
\begin{center}
\psfig{figure=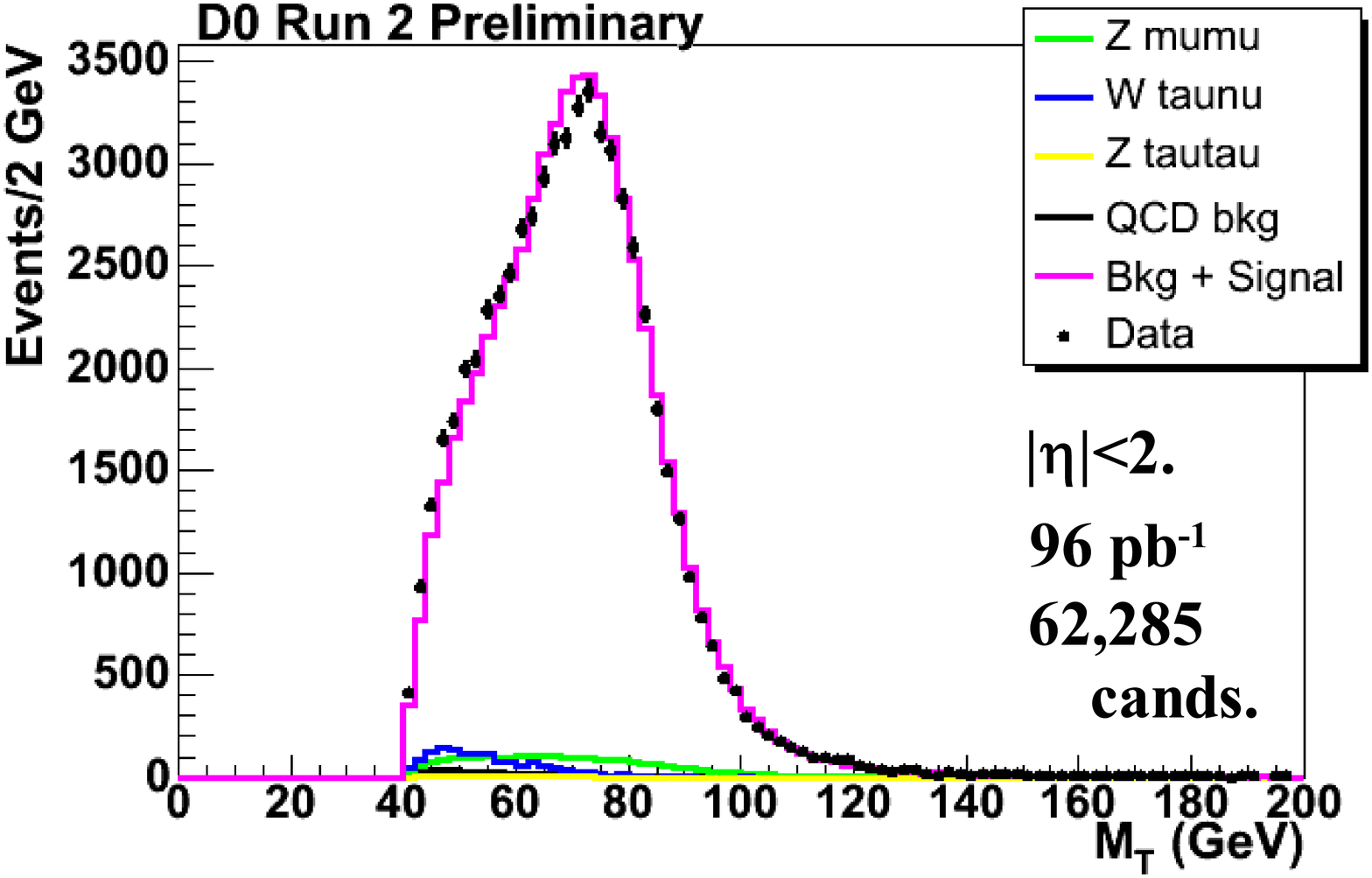,width=6in}
\end{center}
\caption{
The transverse mass of $W \ra \mu \nu$ events from 
96\pb of \dzero\ data\protect\cite{D0:Wmunu}, 
with $|\eta|$ $<$ 2.
}
\label{fig:d0:Wmunu}
\end{figure}

\subsection{$W \ra \tau \nu$ Cross Section}\label{subsec:Wtaunu}

\begin{figure}[ht]
\begin{center}
\psfig{figure=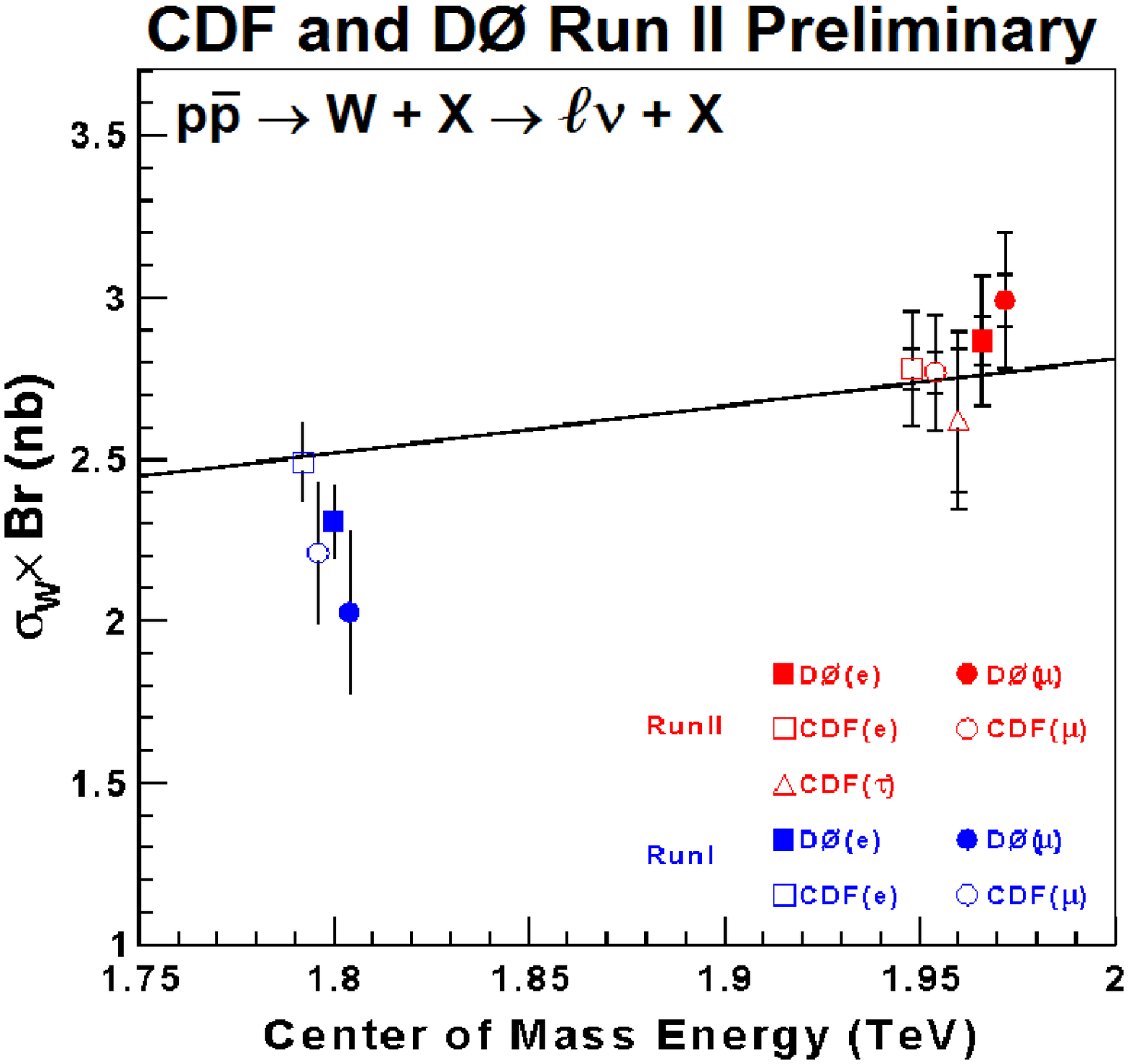,width=4.6in}
\end{center}
\caption{
Summary of W boson
cross section measurements\protect\cite{D0:Wmunu}
from \dzero\ and CDF. Data was taken at 1.8\tev\ or
1.96\tev\ - points are spread for ease of reading.
The line is a Standard Model
prediction\protect\cite{Hamberg:1990np}.
}
\label{fig:Wxsec}
\end{figure}

To identify $W \ra \tau \nu$ events, CDF requires the $E_{T}$ of the
tau to be greater than 25\gev\ and a corresponding \met\ of
25\gev. The largest uncertainty is from tau identification at about 6\% and
the largest backgrounds are from QCD dijets (about 15\%) and 
$W \ra e \nu$ decays
(about 4\%). The cross section calculated from 72\pb of 
data\cite{Safonov:2004zv} 
is given in (\ref{eq:CDF:Wtn}). Leptonic W cross section results are compared 
in Fig.~\ref{fig:Wxsec}.
\dzero\ is working on the $W \ra \tau \nu$ cross section for Run II data.
\be
CDF: \sigma \times \BR{$W \ra \tau \nu$}\, =\,
2620 \pm 70_{stat} \pm 210_{sys} \pm 160_{lum} pb
\label{eq:CDF:Wtn}
\ee

\section{Drell-Yan+Z forward/backward Asymmetry}\label{sec:DYAsymm}

\begin{figure}[ht]
\begin{center}
\psfig{figure=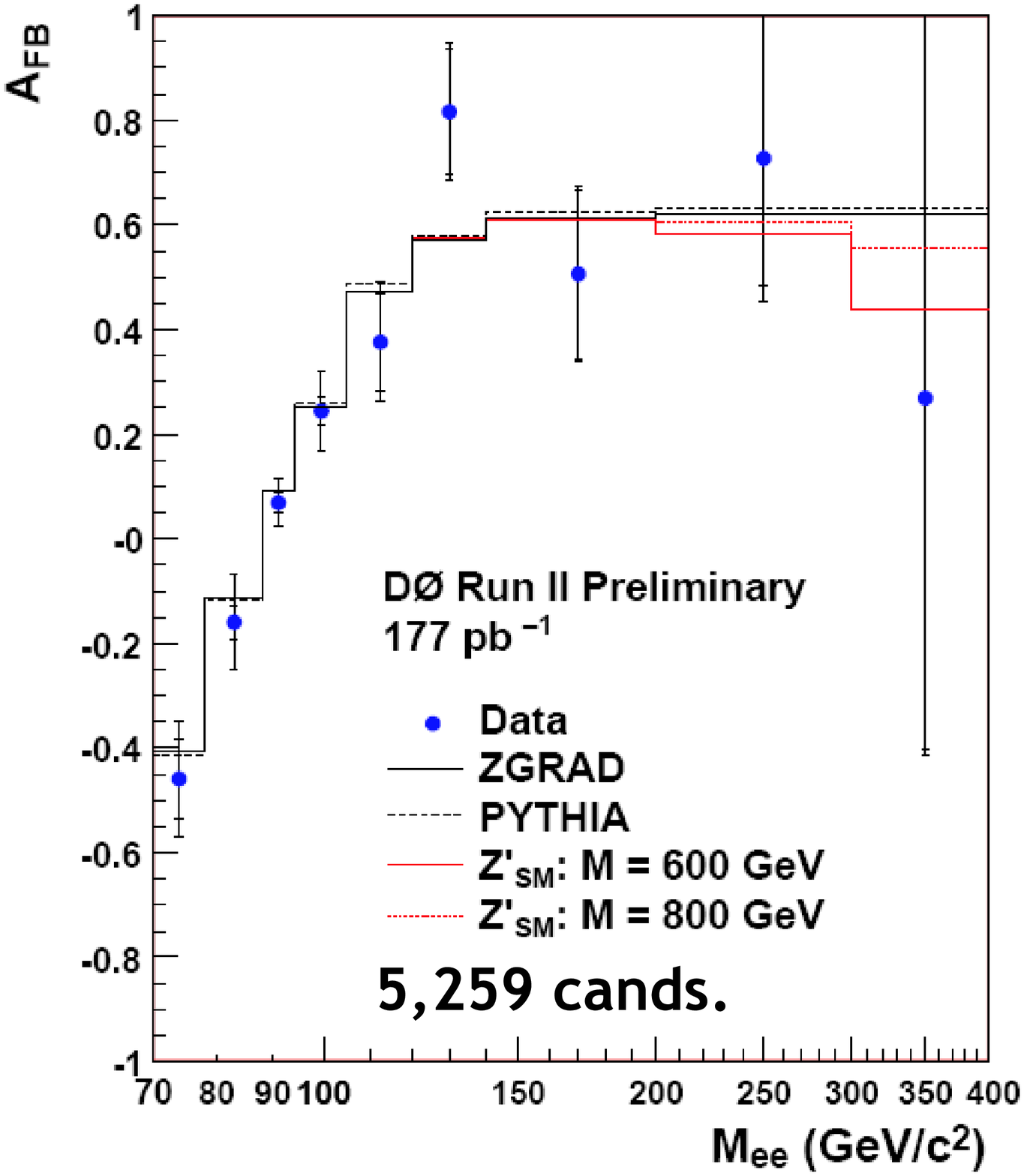,height=6in}
\end{center}
\caption{
The $Z \ra ee$ forward-backward
asymmetry in 177$pb^{-1}$ of \dzero\ Run II data\protect\cite{D0:Afb}.
}
\label{fig:Zasymm}
\end{figure}

One can further test SM predictions by looking at the difference in
lepton production in the proton direction verses that in the 
antiproton direction.
The SM vector and axial-vector couplings of quarks and leptons to 
Z bosons and virtual photons predicts
an asymmetry in the forward and backward cross sections 
($A_{fb}$) of the process
$q\qbar \ra Z/\gamma^{*} \ra ee$ versus the two electron invariant mass. 
See CDF's published article for the Drell-Yan+Z $A_{fb}$
analysis of their 72\pb data set\cite{Acosta:2004wq}. \dzero\ has
recently approved the analysis of 177\pb of data\cite{D0:Afb}, 
and their $A_{fb}$ result is shown in Fig.~\ref{fig:Zasymm}.

\section{W Charge Asymmetry}\label{sec:ChargeAsymm}

\begin{figure}[ht]
\begin{center}
\psfig{figure=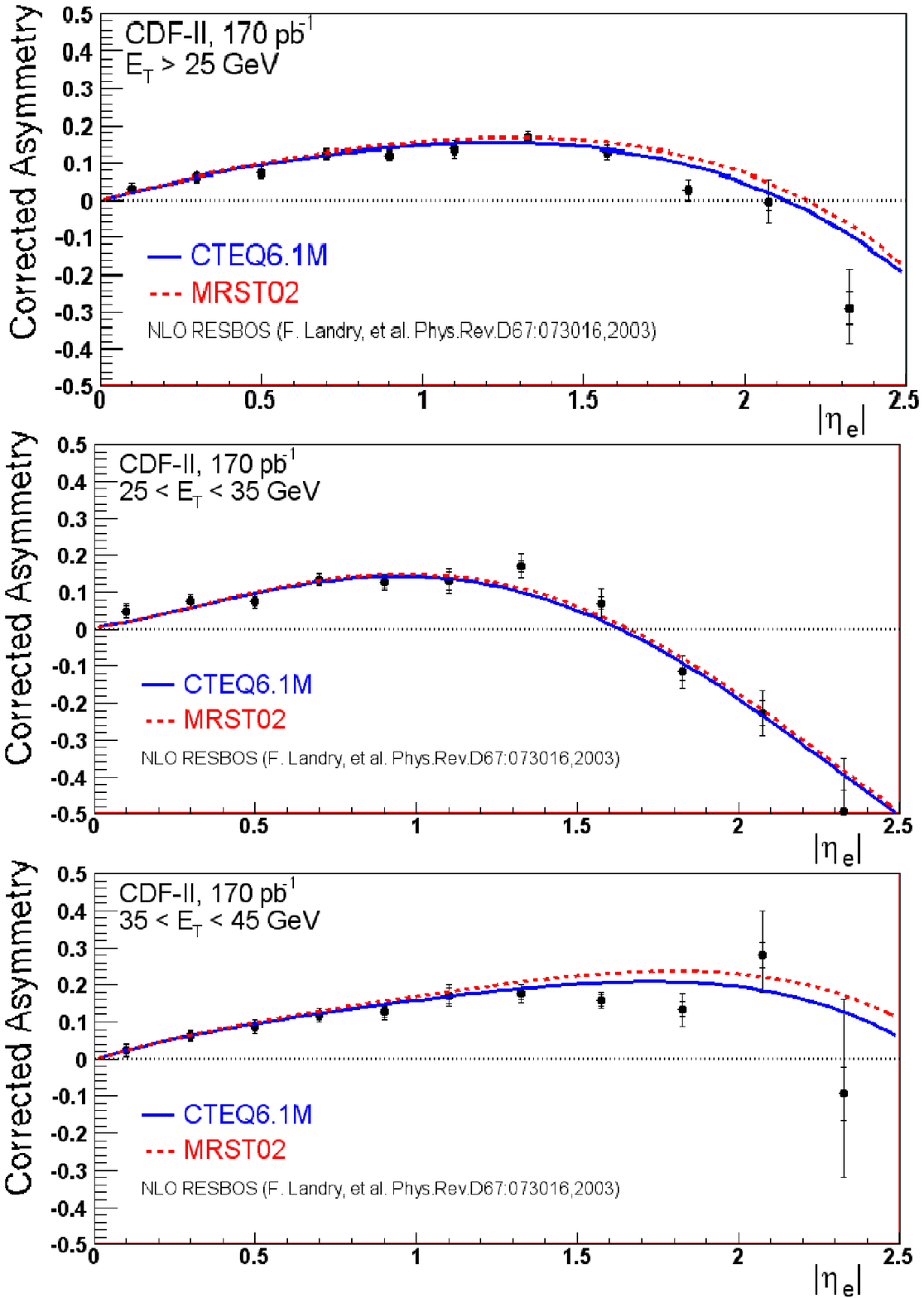,width=5.4in}
\end{center}
\caption{
The W charge asymmetry in 170$pb^{-1}$ of
CDF Run II data\protect\cite{Acosta:2005ud}, broken down by $E_{T}$ range.
}
\label{fig:Wasymm}
\end{figure}

By improving PDFs, systematic uncertainties can be reduced. 
The $u$ quark of the 
proton carries a higher fraction of the particle's 
momentum than the $d$ quark, resulting in $W^{+}$s being boosted in the 
proton direction at hadron colliders. 
Measuring the resulting
$W$ charge asymmetry can be used to improve the $u$ and $d$ PDFs. 
CDF has a new result
for 170\pb of data\cite{Acosta:2005ud}.
Plots of asymmetry versus pseudorapidity for several 
transverse energy ranges are 
shown in Fig.~\ref{fig:Wasymm}.

\section{Summary}\label{sec:Summary}

All results presented agree with SM predictions within errors.
These results also show the \dzero\ and CDF collaborations are
making good progress in measuring electroweak parameter values, gaining better
understanding of their detectors and laying the groundwork
necessary for more complex analyses.


\section*{References}
\begin{thebibliography}{99}

\bibitem{LeCompte:2000st}
T.~LeCompte and H.~T.~Diehl,
Ann.\ Rev.\ Nucl.\ Part.\ Sci.\  {\bf 50}, 71 (2000).

\bibitem{D0:WZe}
The \dzero\ Collaboration,
{\bf \dzero\ Conference Note 4403}, August 12, 2004.

\bibitem{CDF:WZe}
D.~Acosta {\it et al.}  [CDF II Collaboration],
Phys.\ Rev.\ Lett.\  {\bf 94}, 091803 (2005)
[arXiv:hep-ex/0406078].

\bibitem{Hamberg:1990np}
R.~Hamberg, W.~L.~van Neerven and T.~Matsuura,
Nucl.\ Phys.\ B {\bf 359}, 343 (1991)
[Erratum-ibid.\ B {\bf 644}, 403 (2002)].

\bibitem{D0:Zmumu}
The \dzero\ Collaboration,
{\bf \dzero\ Public Note 4573}, August 11, 2004.

\bibitem{Varganov:2004xr}
A.~V.~Varganov,
FERMILAB-THESIS-2004-39

\bibitem{Abazov:2004vd}
V.~M.~Abazov {\it et al.}  [D0 Collaboration],
Phys.\ Rev.\ D {\bf 71}, 072004 (2005)
[arXiv:hep-ex/0412020].

\bibitem{Safonov:2004zv}
A.~Safonov  [the CDF collaboration],
Nucl.\ Phys.\ Proc.\ Suppl.\  {\bf 144}, 323 (2005).

\bibitem{Zsum}
The \dzero\ Collaboration, {\bf \dzero\ Conference Note 4537}, August 11, 2004.

\bibitem{D0:Wmunu}
The \dzero\ Collaboration,
{\bf \dzero\ Note 4750}, March 7, 2005.

\bibitem{Acosta:2004wq}
D.~Acosta {\it et al.}  [CDF Collaboration],
Phys.\ Rev.\ D {\bf 71}, 052002 (2005)
[arXiv:hep-ex/0411059].

\bibitem{D0:Afb}
The \dzero\ Collaboration,
{\bf \dzero\ Conference Note 4757}, March 10, 2005.

\bibitem{Acosta:2005ud}
D.~Acosta {\it et al.}  [CDF Collaboration],
Phys.\ Rev.\ D {\bf 71}, 051104 (2005)
[arXiv:hep-ex/0501023].

\end{thebibliography}
\end{document}